\documentclass[aps,prl,twocolumn,groupedaddress,showpacs]{revtex4-1}

\usepackage{color}
\usepackage{amsmath,amssymb}
\usepackage{graphicx}
\usepackage{bm}
\usepackage{mathrsfs}
\usepackage{textcomp}
\usepackage[normalem]{ulem}
\usepackage{tikz}

\begin{document}

\title{Dynamics of a \emph{Volvox} Embryo Turning Itself Inside Out}%
\author{Stephanie H\"ohn}
\author{Aurelia R. Honerkamp-Smith}
\author{Pierre A. Haas}
\author{Philipp \surname{Khuc Trong}}
\author{Raymond E. Goldstein}%
\affiliation{Department of Applied Mathematics and Theoretical Physics, Centre for Mathematical Sciences, \\ University of Cambridge, 
Wilberforce Road, Cambridge CB3 0WA, United Kingdom}
\date{\today}%
\begin{abstract}
Spherical embryos of the algal genus \emph{Volvox} must turn themselves 
inside out to complete their embryogenesis. This `inversion', which shares important features with morphological events such as 
gastrulation in animals, is perhaps the simplest example of a topological transition in developmental biology. Waves of cell shape changes 
are believed to play a major role in the process, but quantification of the dynamics and formulation of a mathematical 
description of the process have been lacking. Here, we use selective plane illumination microscopy on \emph{V. globator} 
to obtain the first quantitative three-dimensional visualizations of inversion {\it in vivo}. A theory is formulated for inversion 
based on local variations of intrinsic curvature and stretching of an elastic shell, and for the mechanics of an elastic snap-through 
resisted by the surrounding fluid. 
\end{abstract}


\maketitle

Lewis Wolpert's comment, ``It is not birth, marriage, or death, but gastrulation which is truly the most important time 
in your life" \cite{Wolpert}, emphasizes the centrality of topological transitions in developmental biology.  In simple animals, such 
as sea urchin, gastrulation changes the topology of the embryo from that of a simply-connected, hollow sphere of cells to that of a 
torus \cite{McClay1992_seaurchin, Annunziata2014_seaurchin}.   Related deformations of cell sheets involving invagination, but not 
necessarily accompanied by true topological changes, pervade tissue formation in multicellular organisms, where examples include 
neurulation \cite{Lowery2004_neurulation} and ventral furrow formation in {\it Drosophila} \cite{Weischaus_nature}.
One of the common themes is a programme of cell shape changes that result in active bending and stretching of the cell sheet.

Despite intensive study, it has proven difficult to identify model systems in which cells can be tracked with great precision and at the same
time the system is amenable to a simple physical description. Indeed, cell sheet deformations in animal model organisms frequently 
involve additional cellular changes such as division and migration that mask the specific role of ongoing cell shape changes. By contrast, the 
green flagellated alga \emph{Volvox}~\cite{Kirkbook} provides an elegantly simple system to study the dynamic morphology of cell 
sheets: \emph{Volvox} embryos consist of a spherical cellular monolayer which, once cell division is completed, turns itself inside out to bring 
the side of the cell sheet whence emanate the flagella to the outside, and thus enable motility \cite{Viamontes_Kirk_1977, Kirk_Nishii_2001}.

\begin{figure*}
 \includegraphics[width=2.0\columnwidth]{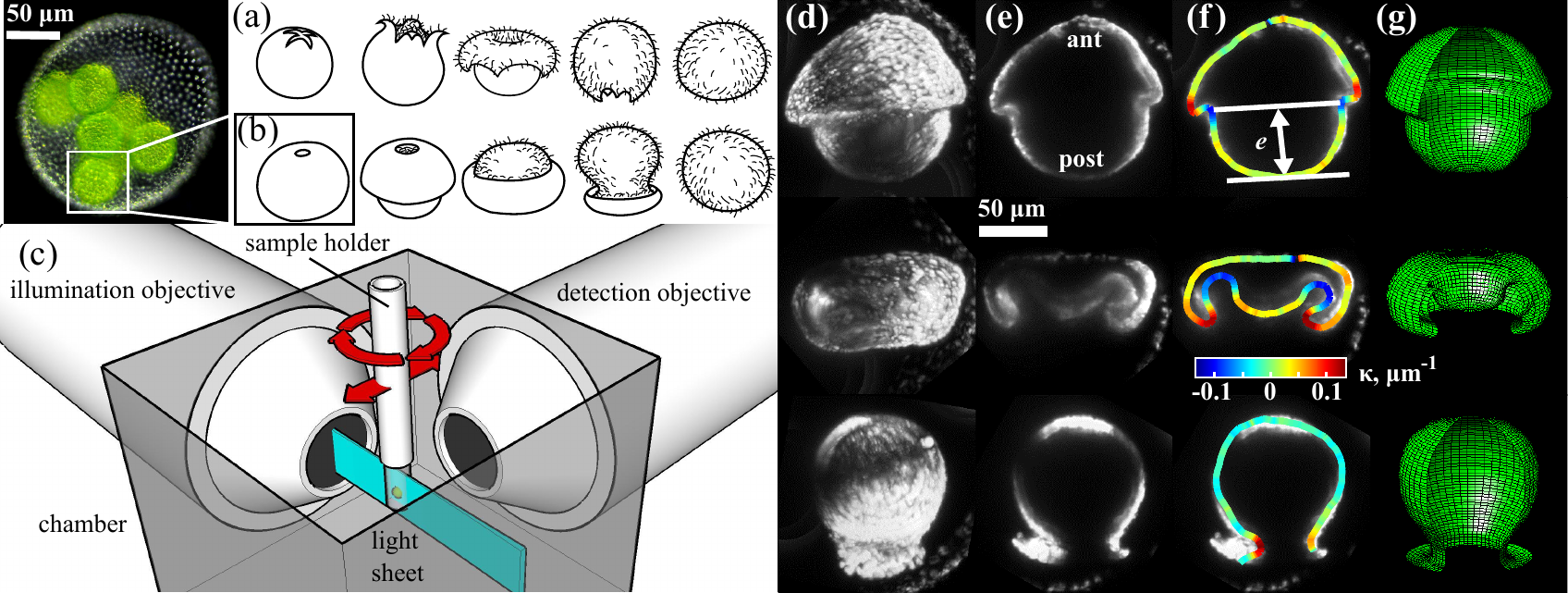}
\caption{(color online).  Inversion.  At top left, an adult spheroid of \emph{V. globator} containing multiple embryos. Species such 
as \emph{V. carteri} perform `type-A' inversion (a), while embryos of \emph{V. globator} use 
`type-B' inversion (b) in order to turn the flagellated surface of the cell sheet to the outside. 
(c) Essential features of the 
 SPIM, in which perpendicular water-dipping objectives are positioned with coincident focal points within the sample chamber.  
(d) A 3D reconstruction of an embryo (autofluorescence) at three 
successive times during inversion.  (e) 
 Optical sections at the midplane of each embryo, 
on which the position of the cell sheet was traced with anterior (`ant') and posterior (`post') poles labelled. 
(f) Traced profiles, with color-coded curvature $\kappa$ and posterior section height $e$ indicated 
on the top profile. (g) Surfaces of revolution obtained from the average contours.}
\label{expfig}
\end{figure*}

The sequence of deformations that \emph{Volvox} embryos undergo during this inversion varies from species to species and is broadly 
divided into two types: in type-A inversion \cite{Viamontes_Kirk_1977, Kirk_Nishii_2001, Hallmann2006}, four outward-curling 
lips open at the anterior pole of the embryo, and peel back to achieve inversion (Fig.~\ref{expfig}a); in type-B inversion 
\cite{Hallmann2006, Hohn2011}, starting with a circular invagination at the equator, the posterior hemisphere inverts, upon 
which an opening widens at the anterior pole, which then peels over the already inverted posterior (Fig.~\ref{expfig}b and Video 1 in the 
Supplemental Material \cite{video1}). In 
particular, type-B inversion shares the feature of invagination with the aforementioned examples of cell sheet deformations 
in animals; this feature is absent in type-A inversion, which starts with the outward folding of a free edge. 
The duration of inversion in different \emph{Volvox} species ranges from 45 minutes to more than one hour. The cells of the 
embryonic cell sheet are connected by a network of cytoplasmic bridges, cell-cell connections resulting from incomplete 
cytokinesis \cite{Green_1981}.

Previous studies \cite{Viamontes_Kirk_1977,Viamontes79,Green_1981,Nishii1999,Nishii2003,Hohn2011} revealed waves of various 
consecutive cell shape changes in combination with cell movements relative to the network of cytoplasmic bridges as mechanisms 
that drive both inversion types. However, the majority of this work used scanning and transmission electron microscopy 
to acquire high-resolution snapshots of embryos arrested and chemically fixed at a particular stage of inversion. 
A single study \cite{Viamontes79} attempted to 
analyze the dynamics of type-A inversion using transmitted light microscopy, putting forth the hypothesis of an elastic snap-through. 
No such quantifications of type-B inversion have yet been reported.  Whereas {\it open} elastic {\it filaments} can
somewhat easily adopt shapes in which the observed local curvature is everywhere equal to the intrinsic curvature, the situation for
elastic {\it sheets} is much more highly constrained and there has been no mathematical study of what we term the 
{\it intrinsic curvature hypothesis}, the conjecture that
local variations in intrinsic curvature can indeed drive the inversion process.  
Here,  using embryos of \emph{V. globator}, we report the first three-dimensional time-lapse visualizations which 
allow quantification of type-B inversion dynamics, and we
propose
the first mathematical descriptions of specific stages of this process. Based on the combination of theory and 
experiment we explore the interplay between active 
deformations and passive responses of the cell sheet and infer its elastic properties.

A selective plane illumination microscope (SPIM) was assembled as previously described~\cite{OpenSPIM}, with minor 
modifications to accommodate an alternate laser ($473$ nm, $144$ mW, Extreme Lasers, Houston, TX) and camera 
(CoolSNAP MYO, $1940\times 1460$ pixels; Photometrics, AZ, USA), and using autofluorescence ($\lambda>$ 500 nm) of the algae for imaging. 
The wild-type {\it Volvox globator} Linn\'{e} strain (SAG 199.80) was obtained from the Culture Collection of Algae at the University of 
G{\"o}ttingen, Germany~\cite{Schlosser_BotActa}. 
Alga cultures were grown axenically in Standard Volvox Medium (SVM)~\cite{Provasoli_Pymatuning} with sterile air bubbling 
in growth chambers (Binder, Germany) set to a cycle of 16 h light ($100\,\mathrm{\mu Em}^{-2}\mathrm{s}^{-1}$, Fluora, OSRAM) at 
$24\,^{\circ}{\rm C}$ and 8 h dark at 
$22\,^{\circ}{\rm C}$. Mother spheroids containing 5-9 embryos at the end of their cell division stage were embedded in 
$1\%$ low melting temperature agarose. The SPIM chamber was filled with SVM at room temperature ($\sim 23\,^{\circ}{\rm C}$) and the 
agarose column was suspended in the laser sheet (Fig.~\ref{expfig}c).  Image stacks were built up by translating the sample perpendicular 
to the observation objective, and were recorded at intervals of 20 to 300 seconds over 2-4 hours to capture inversion of all embryos.  

Where necessary, the datasets were re-sliced to obtain mid-sagittal cross-sections using \textsc{Amira} (FEI, OR, USA). The position 
of the cell sheet was recorded by manually tracing its outline using \textsc{ImageJ} \cite{Rasband_ImageJ} 
on three slices within a $17$ $\mu$m thick region at the 
embryo midplane (Fig.~\ref{expfig}e). The resulting set of points was fitted to a spline and this curve yielded the 
signed curvature $\kappa$ (Fig.~\ref{expfig}f). 
To compute the surface area (Fig.~\ref{expfig}g) of the embryo during inversion, we averaged the shape 
at every time point by reflecting one half of the embryo in its symmetry axis. The latter was determined either by minimizing the 
discrete Fr\'echet distance between one half of the embryo and the reflection of the remaining half, or by the eigenvector 
corresponding to the smallest eigenvalue of the principal component system.

\begin{figure*}
\includegraphics[width=0.98\textwidth]{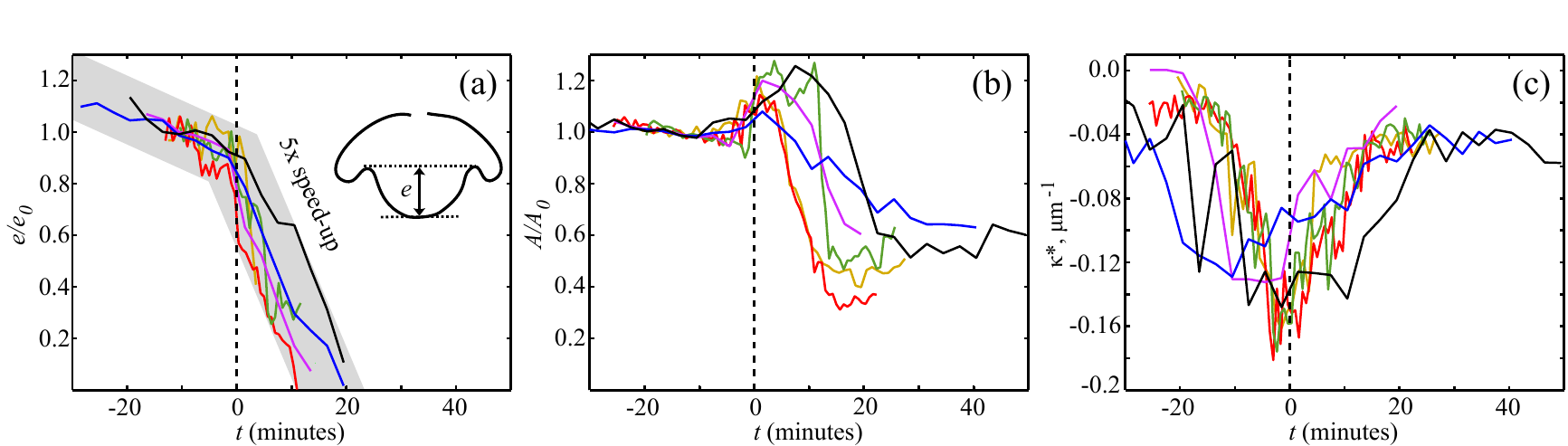}
\caption{(color online).  Dynamic characterization of inversion in \textit{V. globator}. (a) Distance from 
posterior pole to bend region 
decreases at constant speed, which however increases abruptly mid-inversion. Time points for which no bend 
region could be determined are omitted. 
(b) Normalized surface area $A$ of the embryo peaks after the time of speed-up, 
and then decreases.  
(c) Maximum negative embryo curvature $\kappa^*$ peaks concurrently with speed-up; its later relaxation correlates with snap-through. } 
\label{fig:quantification}
\end{figure*}

We shall focus on three key quantities derived from the cross-sectional shapes obtained by the above methods.  Shown 
in Fig. \ref{fig:quantification} for $6$ embryos ranging in initial diameter from $76-130$ $\mu$m, these are 
the distance $e$ from the posterior pole to the bend region (Fig. \ref{expfig}f),  the embryonic
surface area $A$, and the maximum (most negative) value $\kappa^*$ of the curvature in the bend region.
To compare embryos of different size, the posterior distance and surface area were normalized by the values $e_0$ 
and $A_0$ measured in 
the earliest block of time for which data 
exist for all embryos ($\sim -10$ min. and $\sim -20$ min., respectively).

To determine $e$, we first located the posterior pole as the intersection between the embryo shape and the symmetry axis. During early inversion, 
the position of the bend regions was found via the points of largest negative curvature on either side of the symmetry axis. At later stages, 
the points of minimum curvature were used as initial guesses before maximization of the distances along the symmetry axis. The 
posterior-to-bend-distance is then the distance between posterior pole and bend region along the symmetry axis, averaged 
for both embryo halves. Similarly to earlier experimental results for Type-A inversion \cite{Viamontes79}, we find that the 
posterior-to-bend-distance initially decreases at a constant speed (Fig.~\ref{fig:quantification}a), which increases abruptly by 
a factor of about $5$ mid-inversion, reaching values of about $0.03\,\mathrm{\mu m/s}$. We reference all times to the point of speed-up, defined as $t=0$.
An elastic model for this behavior is discussed below.

Previous studies on fixed {\it V. globator} embryos in different inversion stages showed that the most dramatic cell shape change 
event is the formation of wedge-shaped (paddle-shaped \cite{Hohn2011}) cells in the vicinity of the equator. Together with the configuration 
of the cytoplasmic bridges that move to connect these cells at their thin bases, these cell shape changes are believed to impart an 
intrinsic curvature to the cell sheet, driving its observed inward curling.
However, it has been unclear whether active bending causes passive bending or stretching in adjacent regions of the cell sheet.
In Figure \ref{fig:quantification}b, we see that after an initial modest decrease in surface area, 
the surface area first increases significantly to {$110-120\%$} of its initial value, before decreasing to about $55\%$ of 
that value.  The increase in surface area is consistent with the previously reported observation of flattened disc-shaped~\cite{Hohn2011} cells in 
the anterior hemisphere.
The dynamic \emph{in vivo} data allows the different geometric quantities to be related to each other over time. For instance, we see
that the curvature peaks almost concurrently with the speed-up of the posterior (Fig.~\ref{fig:quantification}c), while the surface area 
reaches its maximal value after inversion has sped up (Fig.~\ref{fig:quantification}b). 

In order to formulate a test of the intrinsic curvature hypothesis, and to identify local regions of passive stretching or contraction caused by 
changes in the intrinsic curvature in the equatorial 
invagination region (bend region), we 
consider the axisymmetric deformations of a thin elastic spherical shell of thickness $h$ and undeformed radius $R$ under slow, static variations 
of its intrinsic curvature. (A similar approach, phrased in terms of bending moments rather than preferred curvatures, has been taken 
previously to study the start of involution during gastrulation in \emph{Xenopus laevis} \cite{hardin_xenopus}.)

\begin{figure}[t]
\includegraphics{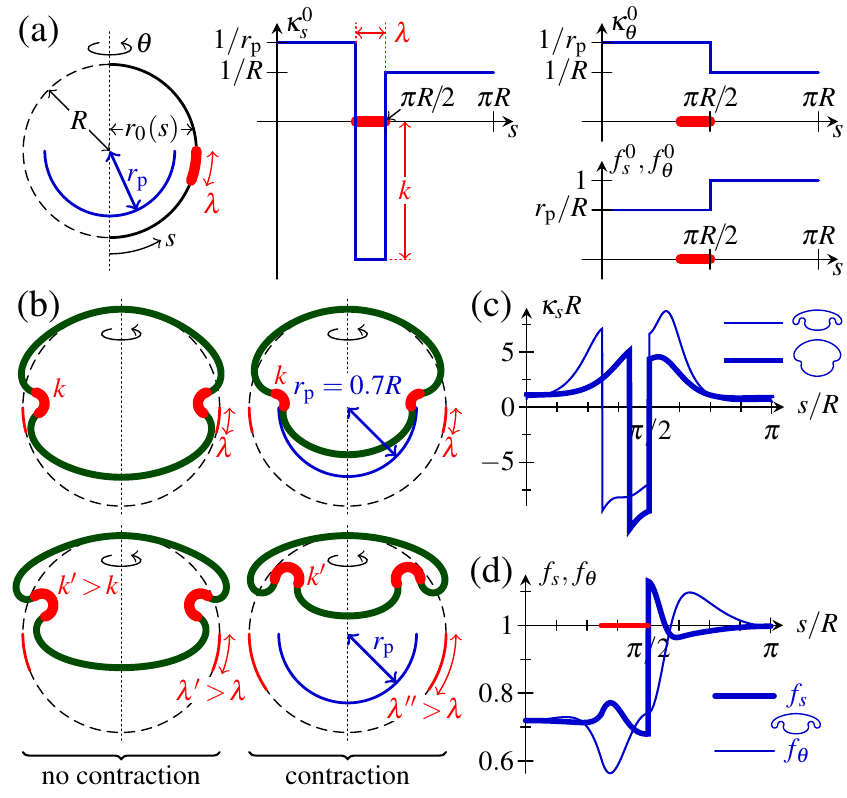} 
\caption{(color online).  Elastic model and results. (a)~Preferred stretches and curvatures at the start of inversion: cell shape changes and 
motion of cytoplasmic bridges impose a negative intrinsic curvature $k$ in a region of width $\lambda$ just below the equator. Posterior 
contraction leads to a preferred posterior radius $r_{\mathrm{p}}<R$. (b)~Numerically obtained cross-sections: in the absence of 
contraction, there is a `purse-string' effect, and the inverted posterior does not fit into the anterior hemisphere. (c)~Meridional curvatures 
for contracted shapes  at early and late invagination stages. (d)~Meridional and circumferential strains at late invagination stage.\label{fig:th1}}
\end{figure}

The distance of the undeformed shell from the axis of revolution is $r_0(s)$, where $s$ is arclength (Fig.~\ref{fig:th1}a); in the deformed 
configuration, this distance is $r(s)$, and arclength is $S(s)$. The meridional and circumferential stretches $f_s=\mathrm{d}S/\mathrm{d}s$ 
and $f_\theta=r/r_0$ define the strains 
\begin{equation}
E_s=f_s-f_s^0,\quad E_\theta=f_\theta-f_\theta^0,
\end{equation}
and curvature strains
\begin{equation}
K_s=f_s\kappa_s-f_s^0\kappa_s^0,\quad K_\theta=f_\theta\kappa_\theta-f_\theta^0\kappa_\theta^0,
\end{equation}
where $\kappa_s$ and $\kappa_\theta$ are the meridional and circumferential curvatures of the deformed shell, respectively. 
As in the Helfrich model \cite{Helfrich} for membranes and generalizations to include `area elasticity' \cite{area_elasticity}, 
$f_s^0,f_\theta^0$ and $\kappa_s^0,\kappa_\theta^0$ introduce preferred stretches 
and curvatures, respectively. Adopting a Hookean model \cite{shelltheory} with elastic modulus $E$ and 
Poisson ratio $\nu$, the deformed shell minimizes the energy
\begin{align}
\mathscr{E}&=\dfrac{2\pi Eh}{1-\nu^2}\int_0^{\pi R}{\hspace{-3mm}r_0\Bigl(E_s^2+E_\theta^2+2\nu E_sE_\theta\Bigr)\,\mathrm{d}s}\nonumber\\
&\hspace{6mm}+ \dfrac{\pi Eh^3}{6(1-\nu^2)}\int_0^{\pi R}{\hspace{-3mm}r_0\Bigl(K_s^2+K_\theta^2+2\nu K_sK_\theta\Bigr)\,\mathrm{d}s}.
\end{align}
In computations we take $\nu=1/2$ and $\varepsilon\equiv h/R=0.15$.

We first consider an initial invagination phase in which the posterior hemisphere starts to invert in response to the formation of region of 
width $\lambda$ of negative preferred curvature $k$ (Fig.~\ref{fig:th1}a). Numerical minimization shows that solely 
imposing a preferred curvature leads to a `purse-string' effect (Fig.~\ref{fig:th1}b) in the vicinity of the equator over large ranges of the 
parameters $k$ and $\lambda$. `Hourglass' shapes of this ilk have been reported as a first stage of inversion in \emph{V. aureus} 
\cite{Kelland77}. Yet, both \emph{V. aureus} and \emph{V. globator} go on to adopt a `mushroom' shape that is not captured by this 
description. Indeed, previous experiments have shown that cells in the posterior hemisphere adopt a thin spindle shape at the start 
of inversion~\cite{Hohn2011}, thereby contracting the posterior hemisphere. To take posterior contraction into account, 
we introduce a reduced preferred posterior radius $r_{\mathrm{p}}<R$. The resulting deformations successfully yield mushroom shapes akin 
to the experimental ones (Fig.~\ref{fig:th1}b), for parameter values $r_{\mathrm{p}}/R\approx 0.7$ and $kR\approx20$. These are in agreement 
with previously published measurements \cite{Hohn2011} of a reduction in cell width in the posterior of about $30-40\%$ and maximal 
radii of curvature in the bend region of $2-3\,\text{\textmu m}$ for embryos of initial radii of $35-45\,\text{\textmu m}$. 

The meridional curvature of the contracted shapes (Fig.~\ref{fig:th1}c) shows an asymmetry between the posterior and anterior hemispheres 
that increases with the progress of invagination. This asymmetry corresponds to the formation of a second bend region, which, in this description, 
arises as a purely passive, material response to the imposed intrinsic curvatures corresponding to the adjacent main bend region.  This result
highlights the point raised earlier, that the local intrinsic curvature can not simply be `read off' from the equilibrium shape.

At a later stage, the contracted posterior fits into the anterior (Fig.~\ref{fig:th1}b), but the uncontracted one does not. Similarly, 
in type-A inversion, embryos in which posterior contraction is prevented biochemically fail to complete inversion \cite{Nishii1999}. All of this 
suggests that contraction is required to create a disparity in the anterior and posterior radii that, in type-B inversion, allows the 
inverted posterior to fit into the as yet uninverted anterior.  This very simple model, with piecewise constant intrinsic curvature
and contraction, produces embryonic shapes in
very good agreement with those found {\it in vivo} (Fig. \ref{expfig}f).

To determine whether or not the disc-shaped cells reported in previous studies \cite{Hohn2011} are the result of 
active shape changes or of passive deformations imposed by forces acting on the cells, we investigate the strains in the theoretical model. 
While the experimental data shows stretching by about $10-20\%$, the model predicts only very slight stretching, of the order of $1\%$. 
Moreover, the model predicts strains within the anterior hemisphere to fall off rapidly (Fig.~\ref{fig:th1}d), so that stretching is confined 
to the bend region. Disc-shaped cells have however been observed throughout the anterior hemisphere \cite{Hohn2011}, hinting at uniform 
stretching. These 
observations suggest that this stretching is the result of active cell shape changes 
rather than of passive deformations. 

\begin{figure}[t]
\includegraphics{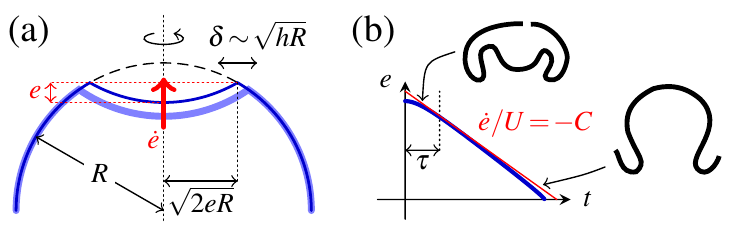} 
\caption{(color online).  Snap-through.  (a) At speed $\dot{e}$, of a dimple of depth $e$ on a thin spherical 
shell of radius $R$ and thickness $h\ll R$. (b)~Numerical solution of~(\ref{eq:dynev}): the snap-through occurs at constant speed after a short transient.\label{fig:th2}}
\end{figure}

Finally, we address the physics underlying the increasing speed of inversion for $t>0$ (Fig.~\ref{fig:quantification}a).  
While the abrupt increase in speed is reminiscent of a mechanical snap-through \cite{Vella14}, conventional snap-throughs of elastic materials 
are associated with phases of acceleration and deceleration, rather than piecewise constant speeds.  
Here, we explore whether such an elastic snap-through, resisted by the fluid surrounding the embryo, agrees with observations. 
To obtain a minimal mathematical description, we consider a spherical shell of radius $R$ and thickness $h$.  
A dimple of depth $e$ pops through at speed $\dot{e}$ in the asymptotic limit $\varepsilon\ll e/R \ll 1$ (Fig.~\ref{fig:th2}a).  
The inverted and non-inverted parts of the shell match up in a narrow asymptotic layer of width $\delta$.  If the fluid has viscosity $\mu$, 
this transition region, of area $\delta (eR)^{1/2}$, gives rise to a hydrodynamic force $\bigl(\mu \dot{e}/\delta\bigr)\delta (eR)^{1/2}$. 
This dominates over the force $\bigl(\mu \dot{e}/R\bigr)eR$ associated with motion of the dimple through the surrounding fluid. 
Hence the dominant hydrodynamic resistance scales as $\mu \dot{e}e^{1/2}R^{1/2}$.

A classical argument \cite{LandauLifshitz} balancing bending and stretching effects in the transition region shows that the elastic driving force of 
the snap-through scales as $Eh^{5/2}e^{1/2}/R$, and yields $\delta\sim\varepsilon^{1/2}R$. Geometry requires $\delta\ll (eR)^{1/2}$, and 
thus $\varepsilon\ll e/R$. Non-dimensionalizing $e$ with $R$ and $\dot{e}$ with the typical scale $U$ of the snap-through speed, the dynamic evolution equation balancing the elastic driving, the hydrodynamic resistance 
and the inertia of the cell sheet (of density $\rho$) takes the form
\begin{equation}
B\Bigl(e\ddot{e}+\tfrac{1}{2}\dot{e}^2\Bigr)=-Ce^{1/2}-\dot{e}e^{1/2}, 
 \label{eq:dynev}
\end{equation}
where $B=\rho Uh/\mu\ll1$ is a Reynolds number and $C=Eh^{5/2}/\mu UR^{3/2}$,
having neglected numerical prefactors. 
An asymptotic analysis of (\ref{eq:dynev}) for $B\ll 1$ reveals that the solution adjusts exponentially to a snapthrough 
at constant speed, $\dot{e}=-C$, in a timescale $\tau\sim RB/U=\rho hR/\mu$. For $\rho\approx10^{3}\,\mathrm{kg/m^3}$, 
$\mu\gtrsim 10^{-3}\,\mathrm{Pa\,s}$ 
(somewhat larger than the viscosity of water), $h\approx 10\,\mathrm{\mu m}$, and $R\approx 40\,\mathrm{\mu m}$, 
we find $\tau\lesssim 10^{-4}\,\mathrm{s}$, essentially instantaneous on the timescale of our measurements, and consistent with the 
kink observed in experiments (Fig.~\ref{fig:quantification}a). Estimating $U\approx 0.03\,\mathrm{\mu m/s}$, we find that
consistency requires $E\gtrsim 2\cdot 10^{-5}\,\mathrm{Pa}$, a very low value, yet 
plausible for a floppy arrangement of cells: 
we estimate the elastic free energy $Eh^{5/2}R^{1/2}\gtrsim 5\cdot 10^{-20}\,\mathrm{J}$, about $10 k_BT$.  

We have shown that the simplest of quasistatic models embodying the intrinsic curvature hypothesis 
is consistent with {\it in vivo} measurements of the early stages of inversion, provided posterior contraction is present.
Hydrodynamic resistance to snapthrough appears to be a consistent explanation for the constant velocity observed, although 
mechanisms such as friction or slow intrinsic dynamics within the cell sheet cannot be ruled out.  
The elastic framework may be extended to address later stages of inversion, including the opening and closure of the phialopore
and the complex dynamics of type-A inversion.  An important open problem is understanding the  
regulation mechanisms that orchestrate the position and timing of cell shape changes that drive inversion.

We are grateful to J. Dunstan and A. Kabla for discussions at an early stage of this work, and to D. Page-Croft and C. Hitch for 
instrument fabrication.  This work was supported in part by an Ernest Oppenheimer Early Career Research Fellowship (ARHS), the EPSRC (PAH), 
and ERC Advanced Investigator Grant 247333 (SH, PKT, and REG).

\end{document}